\documentclass{WileyASNA-v1}

\renewcommand{\vec}{\bm}
\usepackage{bm}
\usepackage{graphicx,xcolor}

\articletype{Original Paper}
\received{}
\revised{}
\accepted{}
\begin{document}

\title{Attempt to obtain the general relativistic planet's motion by special relativity techniques}

\author[1]{Yoshio Kubo*}

\address[1]{\orgdiv{Hydrographic Department (retired), Tokyo, Japan}}

\corres{*Yoshio Kubo, \email{kuboy@sakura.to}}

\presentaddress{5-8-3 Higashi-gotanda, Shinagawa, Tokyo, 141-0022 Japan}

\abstract{It is attempted to derive the general relativistic (GR) equation of motion for planet and its solution solely by the special relativity (SR) techniques.
The motion of a planet relative to the sun and that of the sun to the planet are solved independently in special relativistic framework using the perturbation theory in the celestial mechanics.
The solution reveals a nature of the structure of the spacetime under the gravitation of the sun, and then its effect on the planet's motion is examined.
When the motion thus examined are compared with the one obtained by the general relativity theory in PN approximation, both are different concerning the mean motion and the radius of the orbit but exactly the same as for the perihelion precession.}

\keywords{celestial mechanics, relativity, gravitation}

%%\fundingInfo{Funding info text.}

\maketitle

% \usepackage{newtxtext,newtxmath}

% \usepackage[T1]{fontenc}

% \DeclareRobustCommand{\VAN}[3]{#2}
% \let\VANthebibliography\thebibliography
% \def\thebibliography{\DeclareRobustCommand{\VAN}[3]{##3}\VANthebibliography}

% \usepackage{graphicx}	% Including figure files
% \usepackage{amsmath}	% Advanced maths commands
% \usepackage{amssymb}	% Extra maths symbols

% \renewcommand{\vec}{\bm}
% \usepackage{bm}

\section{Introduction}
\label{S10}

The theory of general relativity gave a crucial solution to the problem of the motion of Mercury's perihelion which had long puzzled the scientists.
Although various theories have been proposed in order to modify or improve Einstein's general relativity theory after that \citep[e.g.]{will, soffel},
his theory is regarded as explaining various physical phenomena sufficiently enough \citep{misner}.

Since the Einstein's spectacular solution, the problem of the precession of Mercury's perihelion has been almost believed
to be solvable only by the theory of general relativity, which is developed based on differential geometry
\citep{roseveare, leverington, bagge}.
In fact, although various attempts were made to solve this problem using only the special theory of relativity,
no solution to give a value close to the observation or to the solution by the general relativity has been obtained \citep{lemmon}.
\citet{bagge} claims to have got a numerical value for the precession coincident with the observation and the general relativity with good accuracy, but its correctness seems dubious.

Most of the efforts attempting to solve the problem of Mercury perihelion in the special relativistic framework are made on the basis of the phenomenon
that the mass of a moving particle changes according to its velocity. 

It is noticed, however, that recently \citet{stepanov} introduces a formula for the perihelion precession
which is derived in the framework of the special relativity and gives the same value as the one obtained by the general relativity.

\citet{kubo} also aims to obtain the planetary motion as close as possible to the observation or to the result by the general relativity, using solely the method of special relativity.
However, the author follows a quite different approach from those researches mentioned above.
In his researches, first the motion of the planet relative to the sun and that of the sun relative to the planet are calculated independently
by the method of the perturbation theory in celestial mechanics, only permitting the Lorentz transformation.
Then, from the obtained results it is concluded that the spatial and time scales in the system fixed to the sun and in the system fixed to the planet are different
and further argued that it is because the spacetime around the sun is not inertial.
Then its structure is examined and shown to be the same as the one obtained by \citet{schwarzschild}. 

In the calculation of the motions of the planet and the sun in \citet{kubo}, however, no satisfactory solution to the motion of the planet's perihelion
or the pericenter in case of the sun's motion (so, written as the pericenter in common usually in the following.) is obtained.
Rather than satisfactory, the amounts of the pericenter motion for the planet and the sun are different; Needless to say, they must be the same.

The present study aims at first to obtain the correct value of the pericenter's precession by performing his calculations again
and this aim is attained successfully.
Then, adding to it, a whole calculation of the relativistic motion of a planet under the spacetime structure introduced above is carried out.
In the process, all the equations of motion, their solutions and so on are approximated to the order of $(v^2/c^2)^1$ with $v$ and $c$ respectively being the velocities of the planet and light.

It means that the present study aims to obtain the equation of motion for a planet and its solution
as a whole that should be equivalent to those by the general relativity theory in the post-Newtonian (PN) or the paramterized PN (PPN) approximation, which are considered to be sufficient enough to discuss the motion of a planet.

The comparison of the present calculation with the solution in PN/PPN approximation shows that
the quantity for the pericenter precession in both solutions are exactly the same
while those of the mean motion and the radius vector are not quite the same.   

\section{Lorentz transformation for three dimensional space}
\label{S20}

As we make full use of the Lorentz transformation throughout the present investigation, let us first examine the transformation formulas a little elaborately.
We introduce the Lorentz transformation for three dimensional space and further pursuit expressions easier to handle.

The formula of the Lorentz transformation for one dimensional space is well-known.
Here we adopt the expression for the transformation found in \citet{goldstein}, which has a form suitable for the extension to the three dimensional formula.
Considering spacetime coordinate systems $(x, y, z, t)$ and $(x', y', z', t')$, both inertial, with
the space in the latter system moving with respect to the former along the respective $z$-axes with the relative velocity $v$,
the transformation of the coordinates between the two systems is given by the following formula:
\begin{equation}
\left(
\begin{array}{c}
x'_1 - x'_0 \\
y'_1 - y'_0 \\
z'_1 - z'_0 \\
ic(t'_1 - t'_0)
\end{array}
\right)
= \textbf{L}_z
\left(
\begin{array}{c}
x_1 - x_0 \\
y_1 - y_0 \\
z_1 - z_0 \\
ic(t_1 - t_0)
\end{array}
\right),
\label{Lorent_z}
\end{equation}
with the Lorentz matrix
\begin{equation}
\textbf{L}_z =
\left(
\begin{array}{cccc}
1 & ~0 &  0& 0 \\
0 & ~1& 0 & 0 \\
0 & ~0 & \gamma & i \beta \gamma \\
0 & ~0 & -i \beta\gamma & \gamma 
\end{array}
\right),
\label{Lz}
\end {equation}
where $\beta = v/c$, with $c$ being the velocity of light, and $\gamma = 1/\sqrt{1 - \beta^2}$.
The subscripts 0 and 1 attached to the coordinates correspond to events 0 and 1, respectively.

The formula in the case where the velocity is not in the direction of $z-$axis but in a general direction represented by $(v_x, v_y, v_z)$
is given as follows \citep{kubo}:
\begin{equation}
\textbf{L} =
\left(
\begin{array}{cccc}
\displaystyle 1 + (\gamma - 1)\frac{v^2_x}{v^2} & \displaystyle (\gamma -1)\frac{v_x v_y}{v^2} & \displaystyle  (\gamma -1)\frac{v_x v_z}{v^2} & \displaystyle i \frac{\gamma}{c}v_x \\
\displaystyle (\gamma -1)\frac{v_y v_x}{v^2} & \displaystyle 1 + (\gamma - 1)\frac{v^2_y}{v^2} &  \displaystyle (\gamma -1)\frac{v_y v_z}{v^2} & \displaystyle i \frac{\gamma}{c}v_y \\
\displaystyle (\gamma -1)\frac{v_z v_x}{v^2} & \displaystyle  (\gamma -1)\frac{v_z v_y}{v^2}  & \displaystyle 1 + (\gamma - 1)\frac{v^2_z}{v^2} & \displaystyle i \frac{\gamma}{c}v_z\\
\displaystyle -i \frac{\gamma}{c}v_x & \displaystyle -i \frac{\gamma}{c}v_y & \displaystyle -i \frac{\gamma}{c}v_z & \gamma 
\end{array}
\right),
\label{Lgen}
\end {equation}
in which $v^2 = v^2_x + v^2_y + v^2_z$ and $\gamma$ is the same as in Eq.~(\ref {Lz}).
The following equation in the same form as Eq.~(\ref{Lorent_z}) holds good as well:
\begin{equation}
\left(
\begin{array}{c}
x'_1 - x'_0 \\
y'_1 - y'_0 \\
z'_1 - z'_0 \\
ic(t'_1 - t'_0)
\end{array}
\right)
= \textbf{L}
\left(
\begin{array}{c}
x_1 - x_0 \\
y_1 - y_0 \\
z_1 - z_0 \\
ic(t_1 - t_0)
\end{array}
\right).
\label{Lorentz}
\end{equation}

We change the shapes of the formula (\ref{Lgen}) and (\ref{Lorentz}) to simpler ones with the use of vector expression.
Keeping the same notation $\textbf{L}$ for the transformation matrix, they come to
\begin{equation}
\textbf{L} =
\left(
\begin{array}{cccc}
\displaystyle \textbf{E} + (\gamma - 1)\frac{\vec{v} \vec{v}^T}{v^2} & ~\displaystyle i \frac{\gamma}{c}\vec{v} \\
\displaystyle -i \frac{\gamma}{c}\vec{v}^T  & \gamma 
\end{array}
\right)
\label{Lgen_v}
\end {equation}
and
\begin{equation}
\left(
\begin{array}{c}
\vec{r}'_1 - \vec{r}'_0 \\
ic(t'_1 - t'_0)
\end{array}
\right)
= \textbf{L}
\left(
\begin{array}{c}
\vec{r}_1 - \vec{r}_0 \\
ic(t_1 - t_0)
\end{array}
\right).
\label{Lorentz_v}
\end{equation}
In these equations, 
\begin{equation}
\textbf{E} = 
\left(
\begin{array}{ccc}
1 & 0 & 0 \\
0 & 1 & 0 \\
0 & 0 & 1 \\
\end{array}
\right),
~~\vec{r} = \left( \begin{array}{c} x \\ y \\ z \end{array} \right),
~~\vec{r}^T = (x, y, z) ~~ \textrm{and so on}.
\end{equation}

Further, if we develop Eq.~(\ref{Lorentz_v}) with Eq.~(\ref{Lgen_v}), we have
\begin{equation}
\begin{split}
\vec{r}'_1 - \vec{r}'_0 &= \vec{r}_1 - \vec{r}_0 + \frac{\vec{v}\left( \vec{v} \cdot (\vec{r}_1 - \vec{r}_0) \right)}{2 c^2}  - \gamma \vec{v} (t_1 - t_0) \\
t'_1 - t'_0 &= -\frac{\gamma}{c^2} \left( \vec{v} \cdot (\vec{r}_1 - \vec{r}_0) \right) + \gamma (t_1 - t_0),
\end{split}
\label{dev1}
\end{equation}
where $(\vec{a} \cdot \vec{b})$ stands for the scalar product of vectors $\vec{a}$ and $\vec{b}$, i.e., $\vec{a}^T \vec{b}$.
The inverse transformation is, 
\begin{equation}
\begin{split}
\vec{r}_1 - \vec{r}_0 &= \vec{r}'_1 - \vec{r}'_0 + \frac{\vec{v} \left(\vec{v} \cdot (\vec{r}'_1 - \vec{r}'_0) \right)}{2 c^2} + \gamma \vec{v} (t'_1 - t'_0) \\
t_1 - t_0 &= \frac{\gamma}{c^2} \left( \vec{v} \cdot (\vec{r}'_1 - \vec{r}'_0) \right) + \gamma (t'_1 - t'_0).
\end{split}
\label{dev2}
\end{equation}
In these equations, the approximation $\gamma - 1 = v^2/(2 c^2)$ is used, which is correct to the order of $(v^2/c^2)^1$.
Therefore, Eqs.~(\ref{dev1}) and (\ref{dev2}) are exact to this order.

\begin{figure}
\centering
\includegraphics[width = 8cm, clip]{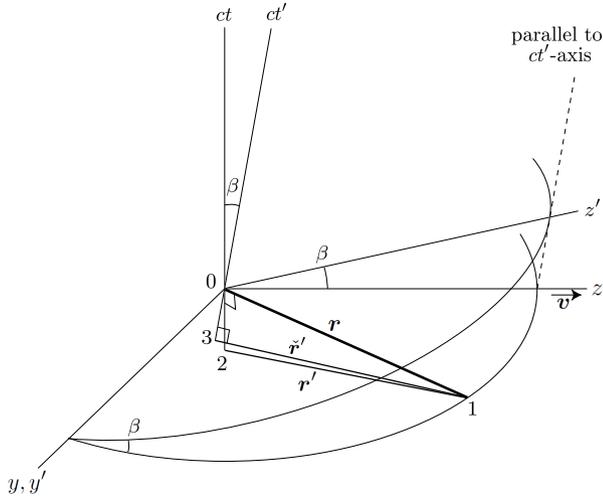}
\caption{Relation between $(x, y, z, t)$ system (X-system) and $(x', y', z', t')$ system (X$'$-system).
They are expressed as $(y, z, t)$ and $(y', z', t')$ coordinates systems, respectively.
As it is difficult to draw the three-dimensional space and time in a figure, the space is represented by $yz$-plane.
The direction of $z$-axis is taken to that of the velocity of X$'$-system with respect to X-system.}
\label{fig1}
\end{figure}

Fig.~\ref{fig1} illustrates the relation between the systems $(x, y, z, t)$ and $(x', y', z', t')$,
which are called X- and X$'$-system, respectively.
Points 0 and 1 are spacetime points or events.
Their coordinates in X-system are $(\vec{r}_i, t_i) ~ (i = 0, 1)$
and those in X$'$-system are $(\vec{r}'_i, t'_i) ~ (i = 0, 1)$,
the quantities all appearing in Eqs.~(\ref{dev1}) and (\ref{dev2}).

The quantity $\vec{r}_i - \vec{r}_j$ stands for the line segment in the space of X-system and it is a space vector if $t_i = t_j$.
It is similar as for X$'$-system.
The quantity $\vec{r}_1 - \vec{r}_0$ is a space vector but $\vec{r}'_1 - \vec{r}'_0$ is not in the ordinary sense because $t'_0 \ne t'_1$.

\section{Relativistic motion of a planet with respect to the sun}
\label{S30}

\subsection{Gravitational force}
\label{S31}

In this section we investigate the relativistic motion of a massless planet around the sun, with the planet moving under the sun's gravitational attraction.

We introduce the  spacetime in which the spatial coordinates of the sun are constant with time, i.e.,
the spacetime fixed to the sun.  We call it S-system.
Similarly, P-system is the spacetime in which the spatial coordinates of the planet are constant with time or the spacetime fixed to the planet.

Besides them we define another spacetime that is coincident with P-system at the moment considered
and moves uniformly to S-system with the planet's instantaneous velocity relative to the sun $\vec{v}$ at that moment.
This spacetime is usually called the instantaneously comoving inertial system but we call it simply P$_0$-system in the present study.
P$_0$-system has no acceleration to S-system.  Of these three systems we treat only S- and P$_0$-systems in this section,
discussing P-system in Section \ref{S40}.

In the following, regard X-system and X$'$-system in Fig.~\ref{fig1}
as S-system and P$_0$-system, respectively.
We consider to obtain the equation of motion for the planet with respect to the sun in S-system.  That is,
if we let Points 0 and 1 in Fig.\ref{fig1} be the positions of the sun and the planet at the time $t = t_0 = t_1$, respectively, we consider the equation of motion at the instant when the sun and the planet are located at these respective points.

We first examine the gravitational force exerted on the planet by the sun in P$_0$-system at the moment 
when the planet is located at Point 1.
Then, it is noticed that the sun is situated at Point 2 at this momont in P$_0$-system. 
If the space coordinates of the sun and the planet at the same instant $t' = t'_1 = t'_2$ are $\vec{r}'_2$ and $\vec{r}'_1$, respectively,
and we put $\vec{r'} = \vec{r}'_1 - \vec{r}'_2$,
the force is given by 
\begin{equation}
\vec{f}' = -\frac{\mu}{{r}'^3} \vec{r}',
\label{f0}
\end{equation}
where $\mu = GM$, with $G$ being the universal gravitational constant and $M$ the rest mass of the sun, respectively.

\subsection{Equation of motion}
\label{S32}

The gravitational force referred to P$_0$-system and given by Eq.~(\ref{f0}) is equal to the acceleration of the planet with respect to the instantaneously comoving inertial frame, that is, the acceleration with respect to P$_0$-system.
That is, we have the following equation of motion in P$_0$-system:
\begin{equation}
\frac{d^2\vec{r}'}{dt'^2}  = \vec{f}' =  -\frac{\mu}{{r}'^3} \vec{r}'. 
\label{alpha0}
\end{equation}

However, the force (\ref{f0}) does not give the acceleration of the planet with respect to the sun in S-system. 
The acceleration in S-system is obtained in the following way, which is not the same as that adopted in \citet{kubo},
although the same result is attained in both ways.

In the formula (\ref{dev2}) consider that events 0 and 1 are for the positions of the planet referred to the sun at some time and at the time a little after that, respectively,
and replace $\vec{r}_1 - \vec{r}_0$ by $d\vec{r}$ and $t_1 - t_0$ by $dt$, and do similarly as for $\vec{r}'$ and $t'$.

Then, Eq.~(\ref{dev2}) becomes
\begin{equation}
\begin{split}
d\vec{r} &= d\vec{r}' + \frac{\vec{v} (\vec{v} \cdot d\vec{r}')}{2c^2} + \gamma \vec{v} dt', \\
dt &= \frac{1}{c^2} ( \vec{v} \cdot d\vec{r}') + \gamma dt',
\end{split}
\label{s31_1}
\end{equation}
where $\vec{v}$ is the velocity of P$_0$-system to S-system and it is equal to that of the planet to the sun at the instant considered.

Dividing the first equation by the second one, we have
\begin{equation}
\begin{split}
\frac{d\vec{r}}{dt}
&=\left[ \frac{d\vec{r}'}{dt'} + \frac{1}{2c^2} \vec{v} \left( \vec{v} \cdot \frac{d\vec{r}'}{dt'} \right) 
+ \gamma \vec{v} \right] \Bigl/
\left[ \frac{1}{c^2} \left( \vec{v} \cdot \frac{d\vec{r}'}{dt'} \right) + \gamma \right] \\
&= \frac{1}{\gamma} \frac{d\vec{r}'}{dt'} - \frac{1}{2c^2}\vec{v} \left( \vec{v} \cdot \frac{d\vec{r}'}{dt'} \right)
+ \vec{v} - \frac{1}{c^2}  \left( \vec{v} \cdot \frac{d\vec{r}'}{dt'} \right) \frac{d\vec{r}'}{dt'}.
\end{split}
\label{s32_2}
\end{equation}
In this equation, $d\vec{r}/dt$ and $d\vec{\vec{r}}'/dt'$ are the velocities of the planet referred to S-system and P$_0$-system, respectively.

By differentiating both sides of Eq.~(\ref{s32_2}) with respect to $t$ we have
\begin{equation}
\begin{split}
\frac{d^2\vec{r}}{dt^2} &= \frac{dt'}{dt} \frac{d}{dt'} \left(\frac{d\vec{r}}{dt} \right) \\
&= \frac{dt'}{dt}\biggl[ \frac{1}{\gamma} \frac{d^2\vec{r}'}{dt'^2} - \frac{1}{2c^2} \vec{v} \left( \vec{v} \cdot
 \frac{d^2\vec{r}'}{dt'^2} \right) \\
&~~~- \frac{1}{c^2} \left( \left( \vec{v} \cdot \frac{d^2\vec{r}'}{dt'^2} \right) \frac{d\vec{r}'}{dt'}
+ \left( \vec{v} \cdot \frac{d\vec{r}'}{dt'} \right) \frac{d^2\vec{r}'}{d^2t'} \right) \biggl]. 
\end{split}
\label{s32_3}
\end{equation}
In this equation, the left-hand side is equal to the acceleration of the planet in S-system.
In the right-hand side we have $d\vec{r}'/dt' = 0$ and as a consequence $dt'/dt = 1/\gamma$ from Eq.~(\ref{s31_1}).
Then, substituting  Eq.~(\ref{f0}) for $d^2 \vec{r}'/dt'^2$, we obtain
\begin{equation}
\frac{d^2\vec{r}}{dt^2} 
= -\frac{\mu}{r'^3} \vec{r}' \left(1- \frac{v^2}{c^2} \right)
+ \frac{\mu \vec{v} (\vec{v} \cdot \vec{r})}{2 c^2 r^3}.
\label{s32_4}
\end{equation}
In deriving this equation, $1/\gamma \cong 1 - v^2/(2 c^2)$ is used as well as $\vec{r}$ and $\vec{r}'$ are not distinguished in the terms with $1/c^2$.

Next we have to express $\vec{r}'$ in the right-hand side of Eq.~(\ref{s32_4}) in terms of the coordinates in S-system.
In the formula (\ref{dev2}), consider here that events 2 and 1 are for the positions of the sun and the planet, respectively, at the time $t' = t'_2 = t'_1$.

Applying the formula to $\vec{r}' = \vec{r}'_1 - \vec{r}'_2$ and taking $t'_1 = t'_2$ as well as $\vec{r}_2 = \vec{r}_0$ into account, we have
\begin{equation}
\vec{r} = \vec{r}_1 - \vec{r}_0 = \vec{r}_1 - \vec{r}_2 = \vec{r}'_1  - \vec{r}'_2 + \frac{\vec{v}(\vec{v} \cdot \vec{r})}{2c^2}
=\vec{r}' + \frac{\vec{v}(\vec{v} \cdot \vec{r})}{2c^2},
\label{s32_5}
\end{equation}
from which it follows
\begin{equation}
\frac{\vec{r}'}{r'^3} = \frac{\vec{r}}{r^3} - \frac{\vec{v}(\vec{v} \cdot \vec{r})}{2c^2r^3} + \frac{3(\vec{v} \cdot \vec{r})^2}{2c^2r^5} \vec{r},
\label{r_r'}
\end{equation}
and from this, we finally obtain the equation of motion
\begin{equation}
\frac{d^2\vec{r}}{dt^2} = -\frac{\mu}{r^3} \vec{r} \left(1- \frac{v^2}{c^2} \right)
+ \frac{\mu (\vec{v} \cdot \vec{r})}{c^2 r^3}\vec{v} - \frac{3\mu (\vec{v} \cdot \vec{r})^2}{2c^2r^5} \vec{r} . 
\label{s32_8}
\end{equation}
This is the same as the corresponding equation in \citet{kubo}.

\subsection{Solution to the equation of motion}
\label{S33}

The equation of motion (\ref{s32_8}) is solved following the Gauss method in celestial mechanics \citep{brouwer}.  More detailed derivation of the solution is found in \citet{kubo}.
First, since the acceleration in Eq.~(\ref{s32_8}) is contained in the plane formed by the vectors $\vec{r}$ and $\vec{v}$,
the motion occurs in this plane.
The acceleration owing to the perturbation is resolved into two components, one along the direction of the radius vector and the other in the direction perpendicular to it on the plane,
being written as $R$ and $S$, respectively.

Then, the time derivatives of the Kepler elements $a, e, \varpi$ and $\epsilon'$ of the planet's orbital motion are calculated,
where $a$ is the semi-major axis, $e$ is the eccentricity, $\varpi$ is the longitude of the perihelion and   
$\epsilon'$ is such an angle as the mean anomaly $l$ is given by
\begin{equation}
 l = \int n dt + \epsilon' - \varpi,
\label{mean_anom}
\end{equation}
with $n ~(= \sqrt{\mu/a^3})$ being the mean motion.
The Gauss formula gives the respective time derivatives of the four orbital elements as follows:
\begin{equation}
\begin{split}
\frac{da}{dt} &= \frac{2}{n \sqrt{1 - e^2}} \left( R e \sin f + S \frac{p}{r} \right), \\
\frac{de}{dt} &= \frac{\sqrt{1 - e^2}}{na}[R \sin f + S (\cos u + \cos f)], \\
\frac{d\varpi}{dt} &= \frac{\sqrt{1 - e^2}}{nae} \left[-R \cos f + S \left(\frac{r}{p} + 1 \right) \sin f \right], \\
\frac{d\epsilon'}{dt} &= - \frac{2r}{na^2}R + \frac{e^2}{1 + \sqrt{1 - e^2}} \frac{d\varpi}{dt} \\
&= - \frac{2r}{na^2}R +(1 - \sqrt{1 - e^2}) \frac{d\varpi}{dt}.
\end{split}
\label{Gauss}
\end{equation}

From the perturbing force in the right-hand side of Eq.~(\ref{s32_8}), the components of the additive acceleration owing to the relativistic effect are as follows:
\begin{equation}
\begin{split}
R &= \frac{\mu \dot {r}^2}{2 c^2 r^2} + \frac{\mu \dot{\theta}^2}{c^2} = \frac{\mu ^3 e^2}{2c^2 r^2 h^2} \sin ^2 f + \frac{\mu h^2}{c^2 r^4}, \\
S &= \frac{\mu \dot{r} \dot{\theta}}{c^2 r} = \frac{\mu ^2 e}{c^2 r^3} \sin f.
\label{s33_compo}
\end{split}
\end{equation}
Putting these quantities into Eq.~(\ref{Gauss}), we have the following explicit equations:
\begin{equation}
\begin{split}
\frac{da}{dt} &= \frac{2a^2}{c^2 h} \left( \frac{\mu^3 e^3}{2r^2 h^2} \sin^3 f + \frac{2\mu e h^2}{r^4} \sin f \right), \\
\frac{de}{dt} &= \frac{h}{c^2 \mu} \biggl( \frac{\mu^3 e^2}{2r^2 h^2} \sin^3 f + \frac{\mu h^2}{r^4} \sin f \\
&~~~+  \frac{\mu^2 e}{r^3} \sin f (\cos u + \cos f) \biggl), \\
\frac{d\varpi}{dt} &= \frac{h}{ c^2 \mu e} \biggl(-\frac{\mu^3 e^2}{2h^2 r^2} \sin^2 f \cos f
- \frac{\mu h^2}{r^4} \cos f \\
&~~~+ \frac{\mu^2 e}{r^2} \left( \frac{1}{r} + \frac{\mu}{h^2} \right) \sin ^2 f \biggl), \\
\frac{d\epsilon'}{dt} &= - \frac{2na}{c^2 \mu} \left( \frac{\mu^3 e^2}{2h^2 r} \sin^2 f + \frac{\mu h^2}{r^3} \right)
+ (1 - \sqrt{1 - e^2}) \frac{d\varpi}{dt}.
\end{split}
\label{Gauss_f}
\end{equation}

Then, integrating the equations, we obtain the changes in the Kepler elements.
In the following, only the main terms, i.e., the lower order terms in the power of the eccentricity $e$, are shown for each element:
\begin{equation}
\begin{split}
\Delta a &= -\frac{\mu}{c^2} \left( 4e \cos f + 2 e^2 \cos 2f + O(e^3) \right), \\
\Delta e &= -\frac{\mu ^2}{c^2 h^2} \left( \cos f + e \cos 2f + O(e^2) \right), \\
\Delta \varpi &= -\frac{\mu ^2}{c^2 h^2} \left( \frac{1}{e} \sin f + \sin 2f + O(e^1) \right), \\
\Delta \epsilon ' &= -\frac{\mu^2}{c^2 h^2} \left( 2f + 2e \sin f + O(e^2) \right)
+ (1 - \sqrt{1 - e^2}) \Delta\varpi,
\end{split}
\label{Gauss_i}
\end{equation}
where $\Delta$ stands for the change in the variables owing to the relativistic effect and $O(e^i)$ the terms with the order of $e^i$ and higher.

From Eq.~(\ref{Gauss_f}) we see that there are no secular terms in $\Delta a$ and $\Delta e$.
That is, both $\langle \Delta a \rangle$ and $\langle \Delta e \rangle$ are zero, with $\langle \cdot \cdot \rangle$ standing for the time average of the inside variable.
As for $\varpi$ we have
\begin{equation}
\begin{split}
\Delta &\langle \frac{d\varpi}{dt} \rangle = \frac{1}{T} \int^T _0 \left( \frac{d\varpi}{dt} \right) dt
= \frac{1}{T} \int^{2\pi} _0 \left( \frac{d\varpi}{dt} \right) \frac{r^2}{h} df \\
&= \frac{\mu^2}{h^2} \frac{1}{T} \int^{2\pi} _0 \biggl( -\frac{\cos f}{e} -2 \cos^2 f + 2 \sin^2 f \\
~~~~~~~  &+ \frac{e}{2} \sin^2 f \cos f - e \cos^3 f \biggl) df = 0,
\end{split}
\label{s33_4}
\end{equation}
rigorously concerning the power of $e$.
Eq.~(\ref{s33_4}) shows that the change in the pericenter motion of the planet caused by the relativistic effect is zero.
By the way, $\Delta \langle d\epsilon '/dt \rangle $ is not zero.

\subsection{Perturbations in the radius vector and the period of the revolution}
\label{S34}

According to the calculation in the previous subsection, there is no change other than periodic one in the semi-major axis $a$ of the planet's orbit.
Nevertheless, a constant increment occurs in the radius vector $r$.

If $r$ is regarded as a function of $a, e$ and $l$, then $\Delta r$ is given by the following equation \citep{kinoshita}:
\begin{equation}
\begin{split}
\Delta r &= \frac{\partial r}{\partial a} \Delta a +  \frac{\partial r}{\partial e} \Delta e +  \frac{\partial r}{\partial l} \Delta l \\
&=\frac{r}{a} \Delta a - a \cos f \Delta e + \frac{ae}{\sqrt{1 - e^2}} \sin f \Delta l.
\end{split}
\label{Del_r}
\end{equation}
For the present we intend to obtain the change in the orbit only for a circular orbit.  
Then, we need $\Delta a$ only to the order of $e^0$ when $a$ is regarded as a power series in $e$.
Also, as for $\Delta e$ the accuracy to the order $e^0$ is necessary as well, and as for $\Delta l$ the accuracy to the order of  $e^{-1}$.
According to this criterion, from Eq.~(\ref{Gauss_i}) we have
\begin{equation}
\begin{split}
\Delta a &= 0, ~~~ \Delta e = -\frac{\mu^2}{c^2 h^2} \cos f + C_e,  \\
\Delta l &= -\frac{3n}{2a} \int \Delta a dt + \Delta  \epsilon ' - \Delta  \varpi = \frac{\mu^2}{c^2 h^2 e} \sin f + C_l.
\end{split}
\label{Del_ael}
\end{equation}
And using these increments in the elements, the following increment in $r$ is obtained:
\begin{equation}
\Delta r = \frac{\mu}{c^2} \left(1 - \cos (f - f_0) \right),
\label{Dr}
\end{equation}
where $f_0$ is the true anomaly at the initial time.
Eq.~(\ref{Dr}) shows that the radius of the perturbed circular orbit is larger than that of the unperturbed orbit by the amount of 
\begin{equation}
\Delta r = \frac{\mu}{c^2},
\label{Dr0}
\end{equation}
with the center of the circle being shifted by the same amount.

Next, the average of the additional change in the time derivative of the planet's longitude $\varpi + f$, which means $\Delta n$, is also calculated as follows: \\
After \citet{kinoshita} we have
\begin{equation}
\Delta f = \frac{a^2}{r^2} \sqrt{1 - e^2} \Delta l + \left( \frac{a}{r} + \frac{1}{1 - e^2} \right) \sin f \Delta e.
\label{Del_f}
\end{equation}
Using this equation and Eqs.~(\ref{Gauss_i}) and (\ref{Del_ael}) we obtain, to the order of $e^0$,
\begin{equation}
\begin{split}
\Delta n &= \langle \frac{d}{dt} \Delta f + \frac{d}{dt} \Delta \varpi \rangle \\
&= \langle -\frac{3n}{2a} \Delta a  +\frac{d}{dt} \Delta \epsilon' + 2\frac{d}{dt}(\sin f \Delta e) \rangle \\
&= -\frac{\mu n a}{c^2 h} \langle 2f \rangle  -\frac{\mu^2}{c^2 h^2} \frac{d}{df}\langle 2 \sin f \cos f \rangle \frac{df}{dt} \\
&= -\frac{2\mu n^2 a}{c^2 h}= -\frac{2\mu^2 n}{c^2 h^2}.
\label{Dn_eq}
\end{split}
\end{equation}
Or, since $nT = 2\pi$, with $T$ being the time duration for one revolution of the planet's orbital motion, Eq.~(\ref{Dn_eq}) may be written as 
\begin{equation}
\Delta T = -\frac{T}{n} \Delta n = \frac{2\mu^2 T}{c^2 h^2}.
\label{DT}
\end{equation}

\section{Motion of the sun relative to the planet}
\label{S40}

\subsection{Motion of the sun in P$_0$-system}
\label{S41}

In this section we discuss the motion of the sun relative to the planet
or the motion of the sun referred to P-system.
But before that we examine the motion of the sun in P$_0$-system.

Throughout this section, $\vec{r}$ and $\vec{v}$ stand for the position and the velocity of the sun relative to the planet, respectively,
that is, $\vec{r}$ and $\vec{v}$ in this section are $-\vec{r}$ and $-\vec{v}$ in the other sections, respectively. 
This rule is applied to the expressions in all the systems, i.e., S-, P$_0$- and P-system.

The velocity and the acceleration of the sun referred to P$_0$-system is given by $d \vec{r}'/dt'$ and $d^2 \vec{r}'/dt'^2$, respectively.
Then, we obviously have 
\begin{equation}
\frac {d\vec{r}'}{dt'} = \vec{v} ~~~ \textrm{and} ~~~ \frac{d^2\vec{r}'}{dt'^2} = 0.
\label{s41_1}
\end{equation}

\subsection{Coordinates transformation between mutually acceleration systems}
\label{S42}

The second equation in Eq.~(\ref{s41_1}) is the equation of motion for the sun written in P$_0$-system.
We must now express it in P-system.

P-system has acceleration to P$_0$-system.
For two systems that accelerate to each other, the Lorentz transformation of coordinates does not hold
but we have to use the formula introduced in \citet{zhukov} as well as in \citet{kubo}.
It is explained as follows:

We consider an inertial system $(x, t)$ and an accelerating system $(x', t')$, with the space in each system being one dimensional.
We suppose that both space origins $x = 0$ and $x' = 0$ coincided with each other and their relative velocity $u = 0$ at the initial time $(t = t' = 0)$.
Let the space origin of the latter system be accelerating
with respect to its instantaneously comoving inertial frame with the acceleration $\alpha$.

On these conditions, the transformation of the small length and time duration at any spacetime point is given by 
\begin{equation}
\begin{split}
dx' &= e^{-\frac{\alpha}{c^2} x'} \frac{dx - u dt}{\sqrt{1 - u^2/c^2}}, \\
dt' &= e^{-\frac{\alpha}{c^2} x'} \frac{dt - (u/c^2)dx}{\sqrt{1 - u^2/c^2}},
\end{split}
\label{s41_1_1}
\end{equation}
or by the inverse transformation
\begin{equation}
\begin{split}
dx &= e^{\frac{\alpha}{c^2} x'} \frac{dx' + u dt'}{\sqrt{1 - u^2/c^2}}, \\
dt &= e^{\frac{\alpha}{c^2} x'} \frac{dt' + (u/c^2)dx'}{\sqrt{1 - u^2/c^2}},
\end{split}
\label{s41_1_2}
\end{equation}
where $u$ is the velocity of the accelerating system with respect to the inertial system at the point.

We apply the formula (\ref{s41_1_2}) to the transformation of the coordinates from P$_0$-system (expressed with the mark $'$) to P-system (with the mark $''$).
In carrying it out, it is a problem which of P$_0$-system or P-system is the inertial system and the other the accelerating system.
Here we notice the fact that no force is felt in P-system (at the place of the planet) because it is making a free fall motion
while the gravitational force is felt in S-system and consequently in P$_0$-system \citep{kubo}.

Then, we must conclude that P-system is the inertial system $(x, t)$,
necessarily with the result that P$_0$-system is the accelerating system $(x', t')$.
Concerning this, we will discuss it again later.
 
We replace $x$ by $r$ and write it in vector although the transformation occurs in one dimensional space.
Then, since $\vec{r}''$ corresponds to $x$, $\vec{r}'$ to $x'$ and so on, Eq.~(\ref{s41_1_2}) is written as
\begin{equation}
\begin{split}
d\vec{r}'' &= \gamma' (d\vec{r}' + \vec{u} dt'), \\
dt'' &= \gamma' \left( dt' + \left(\frac{\vec{u}}{c^2} \cdot d\vec{r}' \right) \right),
\end{split}
\label{s42_2}
\end{equation}
where $\gamma' = e^{\frac{\vec{\alpha}}{c^2} \vec{r}'}/\sqrt{1 - u^2/c^2}$.

\subsection{Equation of motion in P-system}
\label{S43}

In applying the formula (\ref{s42_2}) to the motion of the sun relative to the planet,
first we pay attention to that we only have to consider the initial time $t' = t'' = 0$.
Let $\vec{r}'$ and $\vec{r}' + d\vec{r}'$ be the positions of the sun referred to the planet in P$_0$-system at times $t'$ and $t' + dt'$, respectively,
and similarly as for $\vec{r}''$ and $t''$ in P-system.

From Eq.~(\ref{s42_2}) we have
\begin{equation}
\begin{split}
\frac{d\vec{r}''}{dt''} &= \left( \frac{d\vec{r}'}{dt'} + \vec{u} \right) \Bigl/
\left( 1 + \frac{1}{c^2} \left(\vec{u} \cdot \frac{d\vec{r}'}{dt'} \right) \right) \\
&= \frac{d\vec{r}'}{dt'} + \vec{u} - \frac{1}{c^2} \cdot \frac{d\vec{r}'}{dt'} \left( \vec{u} \cdot \frac{d\vec{r}'}{dt'} \right)
 - \frac{1}{c^2} \vec{u} \left( \vec{u} \cdot \frac{d\vec{r}'}{dt'} \right).
\end{split}
\label{s43_1}
\end{equation}
Then, differentiating both sides with respect to $t''$ and taking into account $d\vec{r}'/dt' = \vec{v}$, $d^2\vec{r}'/dt'^2 = 0$ and $\vec{u} = 0$ at $t' = t'' = 0$, we have
\begin{equation}
\frac{d^2\vec{r}''}{dt''^2}
= \frac{d\vec{u}}{dt''} - \frac{1}{c^2} \left( \frac{d\vec{u}}{dt'}  \cdot \vec{v} \right) \vec{v} .
\label{s43_2}
\end{equation}

In Eq.~(\ref{s43_2}), the vector $ \vec{r}''$ is the vector $\vec{r}'_2 - \vec{r}'_1$ shown in Fig. \ref{fig1}.
The quantity $\vec{u} $ is defined as the velocity of the accelerating system with respect to the inertial system, i.e., the velocity of P$_0$-system with respect to P-system in the present case.
It is also the velocity of the spatial original point of P$_0$-system with respect to the planet.

Let Point 3 in Fig.~\ref{fig1} be the point where $t'$-axis, which is the world line of the original point of
P$_0$-system, intersects the spatial plane of S-system for the time $t = t_1$,
and let $\check{\vec{r}}'$ denote $\vec{r}'_3 - \vec{r}'_1$.
Then, $\vec{u} = -d\check{\vec{r}}'/dt' = d\check{\vec{r}}''/dt''$ at $t' = t''= 0$
and, in view of Eq.~(\ref{f0}) we have
\begin{equation}
\frac{d\vec{u}}{dt''} = \frac{d^2\check{\vec{r}}''}{dt''^2}
= -\frac{\mu}{\check{r}''^3} \check{\vec{r}}'',
\label{s43_4}
\end{equation}
and consequently Eq.~(\ref{s43_2}) comes to
\begin{equation}
\frac{d^2\vec{r}''}{dt''^2} = -\frac{\mu}{\check{r}''^3} \check{\vec{r}}'' % /\sqrt{1 - v^2/c^2}
+ \frac{1}{c^2} \left( \frac{\mu}{r^3} \vec{r} \cdot \vec{v} \right) \vec{v}.
\label{s43_5}
\end{equation}

Now, applying Eqs.~(\ref{dev1}) and (\ref{dev2}) to the relation among $\vec{r}$, $\vec{r}'$ and $\check{\vec{r}}'$, we have the following two equations:
\begin{equation}
\begin{split}
\check{\vec{r}}' &= \vec{r}'_1 - \vec{r}'_3 = \vec{r}'_1 - \vec{r}'_0 =  \vec{r}_1 - \vec{r}_0 + \frac{\vec{v} (\vec{v} \cdot \vec{r})}{2c^2}
- \gamma \vec{v}(t_1 - t_0) \\
& ~~ = \vec{r} +  \frac{\vec{v} (\vec{v} \cdot \vec{r})}{2 c^2}, \\
\vec{r} &= \vec{r}_1 - \vec{r}_0 = \vec{r}_1 - \vec{r}_2 = \vec{r}'_1 - \vec{r}'_2 + \frac{\vec{v} (\vec{v} \cdot \vec{r})}{2c^2}
+ \gamma \vec{v}(t'_1 - t'_2) \\
& ~~ = \vec{r}' +  \frac{\vec{v} (\vec{v} \cdot \vec{r})}{2 c^2},
\end{split}
\label{s43_5}
\end{equation}
and from these we obtain $\check{\vec{r}}' = \vec{r}' + \vec{v} (\vec{v} \cdot \vec{r})/c^2$,
with
\begin{equation}
\check{\vec{r}}'' = \vec{r}'' + \frac{\vec{v} (\vec{v} \cdot \vec{r})}{c^2}
\label{s43_5}
\end{equation}
also holding.
Then, similarly to Eq.~(\ref{r_r'}),
\begin{equation}
\frac{\check{\vec{r}}''} {\check{r}''^3} = \frac{\vec{r}''}{r''^3}
+ \frac{\vec{v}(\vec{v} \cdot \vec{r})}{c^2r^3} - \frac{3(\vec{v} \cdot \vec{r})^2}{c^2r^5} \vec{r},
\label{s43_6}
\end{equation}
and putting this into Eq.~(\ref{s43_4}) we finally obtain the equation of motion for the sun relative to the planet,
\begin{equation}
\frac{d^2\vec{r}''}{dt''^2} = -\frac{\mu}{r''^3} \vec{r}''
+ \frac{3\mu (\vec{v} \cdot \vec{r})^2}{c^2r^5} \vec{r} . 
\label{s43_7}
\end{equation}

Eq.~(\ref{s43_7}) is different from the corresponding equation in \citet{kubo} owing to the procedure
in which the vectors $\vec{r}'$ and  $\check{\vec{r}}'$ are distinguished in the present study.
In \citet{kubo} both the vectors are confused and it causes the contradiction that quantities of the pericenter precession for the planet and for the sun are different.

\subsection{Solution}
\label{S44}

We now solve the equation of motion (\ref{s43_7})
following the same procedure of the perturbation theory as we adopted in obtaining the motion of the planet with respect to the sun.

First, from Eq.~(\ref{s43_7}) we have the components of the perturbing force $R$ and $S$ as follows:
\begin{equation}
R = \frac{3 \mu ^3 e^2}{c^2 r^2 h^2} \sin ^2 f ~~~ \textrm{and} ~~~ S = 0.
\label{s44_1} 
\end{equation}
Putting this into Eq.~(\ref{Gauss}), we have
\begin{equation}
\begin{split}
\frac{da}{dt}
&= \frac{6\mu^3 a^2 e^3}{c^2 h^4}\frac{h}{r^2} \left( \frac{3}{4}\sin f - \frac{1}{4} \sin 3f \right), \\
\frac{de}{dt}
&= \frac{6\mu ^2 e^2}{c^2 h^2} \frac{h}{r^2} \left( \frac{3}{8}\sin f - \frac{1}{8} \sin 3f \right), \\
\frac{d\varpi}{dt}
&= -\frac{6 \mu^2 e}{c^2 h^2} \frac{h}{r^2} \left( \frac{1}{8}\cos f - \frac{1}{8} \cos 3f \right), \\
\frac{d\epsilon '}{dt}
&= -\frac{6 \mu na e^2}{c^2 h} \frac{h}{r^2} ~\frac{1}{1 + e \cos f} \left( \frac{1}{2} - \frac{1}{2} \cos 2f \right) \\
&+ (1 - \sqrt{1 - e^2}) \frac{d\varpi}{dt}.
\end{split}
\label{s44_2}
\end{equation}
The integration gives
\begin{equation}
\begin{split}
\Delta a &= -\frac{\mu^3 a^2}{c^2 h^4}\left( \frac{9 e^3}{2}\cos f - \frac{e^3}{2} \cos 3f \right), \\
\Delta e &= -\frac{\mu ^2}{c^2 h^2} \left( \frac{9 e^2}{4}\cos f - \frac{e^2}{4} \cos 3f \right), \\
\Delta \varpi &= -\frac{\mu^2}{c^2 h^2} \left( \frac{3e}{4}\sin f - \frac{e}{4} \sin 3f \right), \\
\Delta \epsilon ' &= -\frac{\mu n a}{c^2 h} \biggl(3 e^2 f - \frac{3 e^2}{2} \sin 2f - \frac{3 e^3}{2} \sin f  \\
& ~~~  -\frac{e^3}{2} \sin 3f + O(e^4) \biggl) + (1 - \sqrt{1 - e^2}) \Delta\varpi.
\end{split}
\label{s44_3}
\end{equation}
From this equation we see that both $\langle \Delta a \rangle$ and $\langle \Delta e \rangle $ are zero.
As for $\varpi$, we have 
\begin{equation}
\langle \Delta \frac{d\varpi}{dt} \rangle = 0.
\label{s44_3_2}
\end{equation}
This is different from the solution in \citet{kubo}, which says that $\langle \Delta d\varpi/dt \rangle = \mu^2 n/(c^2 h^2)$,
thus resulting in that the pericenter motion is the same for the sun and for the planet.

~
Next, following  Subsection \ref{S34}, we obtain the perturbation in the radius vector and the period of revolution.
According to the criterion that the necessary accuracy for $\Delta a$, $\Delta e$ and $\Delta l$ are respectively $e^0, e^0$ and $e^{-1}$,
we have from Eq.~(\ref{s44_3}),
\begin{equation}
\Delta a = 0, ~~~ \Delta e = 0, ~~ \textrm{and} ~~~ \Delta l = 0.
\label{s44_4}
\end{equation}
And from this, obviously we have, to the order of $e^0$,
\begin{equation}
\Delta r = 0.
\label{s44_5}
\end{equation}
Also, after Eq.~(\ref{Dn_eq}) and from Eq.~(\ref{s44_3}), to the order of $e^0$ we have
\begin{equation}
\Delta n = 0, ~~~ \textrm{consequently} ~~~ \Delta T = 0.
\label{s44_6}
\end{equation}

\section{Structure of the spacetime around the sun}
\label{S50}

Let the radius and the period of the revolution in a non-relativistic circular orbit of a planet be $a_0$ and $T_0$, respectively. 
A result from the discussion in Section \ref{S30} is that the relativistic orbit that has started with the same initial conditions
is also a circular orbit but with the radius and the revolution period given by
\begin{equation}
a_1 = \left( 1 + \frac{\mu}{c^2 a} \right) a_0 ~~~ \textrm{and} ~~~ T_1 = \left( 1 + \frac{2\mu}{c^2 a} \right) T_0.
\label{s5_1}
\end {equation}

On the other hand, a result from Section \ref{S40} is that the relativistic orbit of the sun around the planet with the same initial conditions is
also circular and with the radius and the period of the revolution, respectively,
\begin{equation}
a_2 = a_0 ~~~ \textrm{and} ~~~ T_2 = T_0.
\label{s5_2}
\end {equation}
This is a contradiction and the resolution is obtained only by assuming that the spatial and time scales in S-system and P-system are different \citep{kubo}.

Both spatial and time scales in S-system are smaller than those in P-system.
And Eq.~(\ref{s5_2}) seems to suggest that the scales in P-system are the same as in the inertial spacetime.
In fact, no force is felt for an observer in P-system who moves with the planet, which is making a free fall motion,
while for an observer at rest anywhere in S-system the force owing to the sun's gravitation would be felt.
From this, it is argued that the inertial space around the sun, or a central mass generally, is not the space at rest with respect to the mass
but the space falling or escaping (for both motions are possible) with the speed of $\pm \sqrt{2\mu/r}$ and with the acceleration of $-\mu/r^2$ toward/from the mass.
In that case, the planet is regarded as making a free fall motion in this inertial system locally.
We call the spacetime with this falling/escaping space F-system.

So observing, if we express the coordinates in F-system by $(r, \theta, t)$ and those in the sun-fixed system, which we call now $\tilde{\textrm{S}}$-system,
by $(\tilde{r}, \tilde{\theta}, \tilde{t})$,
we have the following relation between them: 
\begin{equation}
\begin{split}
\tilde{r} &= r - \frac{\mu}{c^2} = \left(1 - \frac{\mu}{c^2 r} \right) r, \\
d\tilde{r} &= dr, \\
d\tilde{\theta} &= d\theta, \\
d\tilde{l} &= \tilde{r} d\tilde{\theta} = \left(1 - \frac{\mu}{c^2 r} \right) r d\theta = \left( 1 - \frac{\mu}{c^2 r} \right) dl, \\
d\tilde{t} &= \left( 1 - \frac{2 \mu}{c^2 r} \right) dt,
\end{split}
\label{s5_3}
\end{equation}

Now, we note that the sun is at rest referred to the inertial space existing at the infinite distance from the sun.
The far distant inertial system, which may be called I-system, does not exist near a mass but instead F-system exists there.
However, we can imagine a system that is constructed by extending I-system to everywhere in the universe whether there exist masses or not. 
If we admit such an imaginary system around the sun,
it is nothing else but the spacetime that we have called S-system so far.
Eq.~(\ref{s5_3}) can also be regarded as giving the relation between $\tilde{\textrm{S}}$-system and S-system.
We distinguish the spacetime systems expressed in term of the coordinates with the timescales affected by the mass from those written in terms of the scales in the inertial systems, and attach the mark "~$\tilde{~}$~" to the former systems, such as $\tilde{\textrm{S}}$-system, $\tilde{\textrm{P}}_0$-system and so on.  

\section{Motion of a planet in $\tilde{\textrm{S}}$-system}
\label{S60}

If we look at the results obtained in Sections \ref{S30} and \ref{S40} carefully again,
we first notice that the sun's motion referred to P-system is the same as that in the Newtonian mechanics,
that is, there appears no perturbation owing to the relativistic effects except small periodic ones.

As for the planet's motion referred to $\tilde{\textrm{S}}$-system, the perturbation $\Delta r$ and $\Delta n$ are added to the Newtonian motion,
but they are only apparent perturbations owing to the difference of the spatial and time scales in $\tilde{\textrm{S}}$- and P- systems.
If the scale difference in $\tilde{\textrm{S}}$-system is admitted and suitably taken into account, it is considered that the planet's motion would be the same as the sun's motion.
To sum up these facts, we can say that no virtual perturbation both in the planet's and the sun's motions is detected in the process so far. 

Meanwhile, the difference in the scales in the two systems with and without  the mark "~$\tilde{~}$~" brings about a real perturbation in the motions of the planet and the sun,
and the perturbing force thus produced changes the radius vector $r$ and the mean motion $n$
as well as $\langle d\varpi/dt \rangle$.
The changes are common to the motions of the planet and of the sun.

\subsection{Equation of motion}
\label{S61}

F-system is falling to (or escaping from) the sun with the speed of $\pm \sqrt{2\mu/r}$.
The falling/escaping motion of F-system is regarded as the one referred to S-system,
i.e., we can say that the motion of F-system is the one observed in S-system.

Now, we examine what the motion of F-system looks like when observed in $\tilde{\textrm{S}}$-system.
The motion of F-system with respect to S-system is represented by the motion of any point that costitutes the space of the system
and the equation of motion for the point in S-system is described as
\begin{equation}
\frac{d^2r}{dt^2} = - \frac{\mu}{r^2},
\label{s61_10}
\end{equation}
a one dimensional equation since the motion of F-system is spherically symmetric.

Transforming the variables $(r, t)$ to $(\tilde{r}, \tilde{t})$ using Eq.~(\ref{s5_3}), we have
\begin{equation}
\frac{dr}{dt} = \frac{d\tilde{r}}{d\tilde{t}}  \left( 1 - \frac{2\mu}{c^2r} \right).
\label{s61_2}
\end{equation}
and from this
\begin{equation}
\begin{split}
\frac{d^2 r}{dt^2} &= \frac{d\tilde{t}}{dt} \frac{d}{d\tilde{t}} \left[ \frac{d\tilde{r}}{d\tilde{t}}\left( 1- \frac{2\mu}{c^2 r} \right) \right] \\
&= \left( 1 - \frac{2\mu}{c^2 r} \right)^2 \frac{d^2 \tilde{r}}{d\tilde{t}^2} 
+ \frac{2\mu}{c^2 r^2} \left( \frac{dr}{dt}\right)^2,\\
\end{split}
\label{s61_3}
\end{equation}

Then, using also $\tilde{r} = r (1 - \mu/(c^2 r))$, Eq.~(\ref{s61_10}) becomes
\begin{equation}
\left( 1 - \frac{2\mu}{c^2 r} \right)^2 \frac{d^2 \tilde{r}}{d\tilde{t}^2} 
+ \frac{2\mu}{c^2 r^2} \left( \frac{dr}{dt}\right)^2 = \left( 1 - \frac{\mu}{c^2 r} \right)^2 \frac{\mu}{\tilde{r}^2}.
\label{s61_11}
\end{equation}
and taking into account $(1/c^2) d^2 r/dt^2 = -(1/c^2) \mu/r^2$ and $(dr/dt)^2 = 2\mu/r$, we have
\begin{equation}
\frac{d^2\tilde{r}}{d\tilde{t}^2} = - \frac{\mu}{\tilde{r}^2} - \frac{6\mu^2}{c^2 r^3}.
\label{s61_11}
\end{equation}

Eq.~(\ref{s61_11}) shows that the force $-6\mu^2/c^2 r^3$ appears in addition to the inverse square force of Newtonian gravitational force in the equation of motion in $\tilde{\textrm{S}}$-system,
and this fact demands to modify the equation of motion (\ref{alpha0}) that is written in P$_0$-system
as the following one in $\tilde{\textrm{P}}_0$-system:
\begin{equation}
\frac{d^2\tilde{\vec{r}}'}{d\tilde{t}'^2}  = -\frac{\mu}{{\tilde{r}}'^3} \vec{r}' - \frac{6\mu^2}{c^2 r^4} \vec{r}. 
\label{alpha_t}
\end{equation}
Then, following the same procedure in Subsection \ref{S32}, we have the equation of motion for the planet with respect to the sun in $\tilde{\textrm{S}}$-system, as follows:
\begin{equation}
\frac{d^2\tilde{\vec{r}}}{d\tilde{t}^2} = -\frac{\mu}{\tilde{r}^3} \tilde{\vec{r}} \left(1- \frac{v^2}{c^2} \right)
- \frac{6\mu^2}{c^2 r^4}\vec{r}
+ \frac{\mu (\vec{v} \cdot \vec{r})}{c^2 r^3}\vec{v}
- \frac{3\mu (\vec{v} \cdot \vec{r})^2}{2c^2r^5} \vec{r} . 
\label{s61_13}
\end{equation}
 
\subsection{Solution}
\label{S62}

In this section we omit the $\tilde{~~}$ marks to be attached to the variables for brevity, on understanding that they are the coordinates expressed in $\tilde{\textrm{S}}$-system. 

The perturbation terms in Eq.~(\ref{s61_13}) is the sum of the original ones in Eq.~(\ref{s32_8}) and the additional ones in Eq.~(\ref{alpha_t}).
The latter, i.e., the additional perturbation terms are as follows:
\begin{equation}
R_+ = -\frac{6\mu^2}{c^2 r^3} ~~~ \textrm{and} ~~~ S_+ = 0.
\label{s62_1}
\end{equation}
Then, the additive parts of the time derivatives of the orbital elements owing to Eq.~(\ref{s62_1}) are as follows:
\begin{equation}
\begin{split}
\frac{da}{dt}_+ &=
-\frac{12\mu^3 a^2}{c^2 h^4} \frac{h}{r^2} \left( e \sin f + \frac{e^2}{2} \sin 2f \right), \\
\frac{de}{dt}_+ &=
-\frac{6\mu ^2}{c^2 h^2} \frac{h}{r^2} \left( \sin f + \frac{e}{2} \sin 2f \right), \\
\frac{d\varpi}{dt}_+ &=
\frac{6 \mu^2}{c^2 h^2} \frac{h}{r^2} \left( \frac{1}{2} + \frac{\cos f}{e} + \frac{1}{2} \cos 2f \right) \\
\frac{d\epsilon '}{dt}_+ &=
\frac{12 \mu na}{c^2 h} \frac{h}{r^2} + (1 - \sqrt{1 - e^2}) \frac{d\varpi}{dt}_ +.
\end{split}
\label{s62_2}
\end{equation}
And the integration gives the additional parts to be added to Eq.~(\ref{Gauss_i}), as
\begin{equation}
\begin{split}
\Delta a_+ &= \frac{12\mu^3 a^2}{c^2 h^4}\left(e\cos f + \frac{e^2}{4} \cos 2f \right), \\
\Delta e_+ &= \frac{6\mu ^2}{c^2 h^2} \left( \cos f + \frac{e}{4} \cos 2f \right), \\
\Delta \varpi_+ &=  \frac{6 \mu^2}{c^2 h^2} \left( \frac{f}{2} + \frac{\sin f}{e} + \frac{1}{4} \sin 2f \right), \\
\Delta \epsilon '_+ &= \frac{12 \mu na}{c^2 h} f + (1 - \sqrt{1 - e^2}) \Delta \varpi_+.
\end{split}
\label{s62_3}
\end{equation}
Carrying out the summation of Eq.(\ref{Gauss_i}) and Eq.~(\ref{s62_3}) actually,
the explicit solution for the changes in the orbital elements owing to the relativity is
\begin{equation}
\begin{split}
\Delta a &= \frac{\mu}{c^2} \left( 8e \cos f + e^2 \cos 2f + O(e^3) \right), \\
\Delta e &= \frac{\mu ^2}{c^2 h^2} \left( 5\cos f + \frac{e}{2} \cos 2f + O(e^2) \right), \\
\Delta \varpi &= \frac{\mu ^2}{c^2 h^2} \left( 3f+ \frac{5}{e} \sin f + \frac{1}{2}\sin 2f + O(e^1) \right), \\
\Delta \epsilon ' &= \frac{\mu^2}{c^2 h^2} \left( 10f - 2e \sin f + O(e^2) \right)
+ (1 - \sqrt{1 - e^2}) \Delta\varpi.
\end{split}
\label{s62_4}
\end{equation}
That is, Eq.~(\ref{s62_4}) is the solution to Eq.~(\ref{s61_13}).
 
Summing up the whole solutions obtained so far, we see that there are no secular terms in $\langle \Delta a \rangle$ and $\langle \Delta e \rangle$ in total,
that is, no secular motions appear in $a$ and $e$ owing to the effect of the relativity.

However, similar to the discussion in Subsection \ref{S34}, there exist $\Delta r$ and $\Delta n$.
We calculate them following the discussion in Subsection \ref{S34} almost as it is.

From Eq.~(\ref{s62_4}), we have
\begin{equation}
\Delta a = 0,  ~~~  \Delta e = \frac{5\mu^2}{c^2 h^2} \cos f + C_e ~~~ \textrm{and} ~~~ \Delta l =  -\frac{5\mu^2}{c^2 h^2 e} \sin f + C_l,
\label{s62_5}
\end{equation}
respectively to the order of $e^0$ for $\Delta a$ and $\Delta e$ and  to $e^{-1}$ for $\Delta l$, the same criterion as in Subsection \ref{S34}.

We observe that Eq.~(\ref{Del_r}) shows a proportional relationship
between the ``vector'' $(\Delta a, \Delta e, \Delta l)$ and $\Delta r$.  Thus,
the change in the radius vector $r$ is
\begin{equation}
\Delta r = -\frac{5\mu}{c^2},
\label{s63_5}
\end{equation} 
where it must be emphasized that this value is common to the motion of the planet relative to the sun and that of the sun relative to the planet.

Similarly as for $\Delta n$, it is proportional to the ``vector'' $(\Delta e, \Delta l)$ according to Eq.~(\ref{Del_f}),
and therefore we have the change in the mean motion $n$ is
\begin{equation}
\Delta n = \frac{10\mu^2 n}{c^2 h^2}, ~~ \textrm{or} ~~ \Delta T = -\frac{10\mu^2 T}{c^2 h^2},
\label{s63_6}
\end{equation} 
for both motions of the planet relative to the sun and of the sun relative to the planet.

It must be remembered, however, that the discussion in Subsection \ref{S34} is valid only for a circular orbit,
that is, the terms higher than $e^1$ are neglected.
Eqs.~(\ref{s63_5}) and (\ref{s63_6}) are also correct only to $e^0$ and they contain the errors of $O(e^1)$.
More exact calculation for $\Delta r$ and $\Delta n$ is possible but would be considerably troublesome.

Meanwhile, as for the precession of the pericenter, $\Delta (d\varpi/dt$),  we have from Eqs.~(\ref{s33_4}) and (\ref{s62_3}),
\begin{equation}
\langle \Delta \frac{d\varpi}{dt} \rangle = \langle \Delta \frac{d\varpi}{dt}_+ \rangle = \frac{3 \mu^2 n}{c^2 h^2},
\label{s63_3}
\end{equation}
rigorously concerning the expansion in the powers of $e$.
Again, Eq.~(\ref{s63_3}) is common to both the planet's and the sun's motions.

In Eq.~(\ref{s63_3}), if $h^2/\mu = a(1 - e^2)$ and $n = 2\pi/T$ are substituted, it is written as
\begin{equation}
\langle \Delta \frac{d\varpi}{dt} \rangle = \frac{6\pi \mu}{c^2 a(1 - e^2) T},
\label{s7_1}
\end{equation}
This is equivalent to the Einstein's perihelion formula by his general relativity theory \citep{einstein},
\begin{equation}
\varepsilon = 24 \pi^3 \frac{a^2}{T^2 c^2 (1 - e^2)},
\label{einst}
\end{equation}
where the unit of $\varepsilon$ is the angle (in radian) that the perihelion motion gains during one revolution of the planet.

\section{Comparison with PN/PPN equation of motion}
\label{S70}

It may be expected that Eq.~(\ref{s61_13}) is consistent with the post-Newtonian (PN) equation of motion
for a planet, i.e., the equation of motion by the general relativity theory approximated to the order of $v^2/c^2$, or the parameterized PN (PPN) equation of motion.

For example,  Eq.~(\ref{s61_13}) is to be compared with the equation (3.1.46) in \citet{brumberg}, which reads
\begin{equation}
\begin{split}
\ddot{\vec{r}} + \frac{GM}{r^3}\vec{r} &= \frac{m}{r^3} \biggl[ \biggl( 2( \beta + \gamma - \alpha) \frac{GM}{r}
- (\gamma + \alpha) \dot{\vec{r}}^2 + 3\alpha \frac{(\vec{r} \dot{\vec{r}} )^2}{r^2} \biggl) \vec{r} \\
&+ 2(\gamma + 1  -\alpha) (\vec{r} \dot{\vec{r}} ) \dot{\vec{r}} \biggl].
\end{split}
\label{s70_1}
\end{equation}
with $GM = \mu$ and $m = \mu/c^2$.
Also, $\beta$ and $\gamma$ are the parameters with the values of 1 for both in the general relativity
while $\alpha$ is the parameter in the PPN equation whose value depends on the adopted model.

Eq.~(\ref{s70_1}) is rewritten in our style as
\begin{equation}
\begin{split}
\frac{d^2\vec{r}}{dt^2} =
-\frac{\mu}{r^3} \vec{r} \left(1+ (\gamma + \alpha)\frac{v^2}{c^2} \right)
+ 2(\beta + \gamma - \alpha) \frac{\mu^2}{c^2 r^4}\vec{r} \\
+ \frac{3\alpha \mu (\vec{v} \cdot \vec{r})^2}{c^2r^5} \vec{r}
+ \frac{2(\gamma + 1 - \alpha)\mu (\vec{v} \cdot \vec{r})}{c^2 r^3}\vec{v}.
\end{split}
\label{s70_2}
\end{equation}
When we compare  Eq.~(\ref{s70_2}) and Eq.~(\ref{s61_13}), we see they are quite different.
Adopting $\beta = \gamma = 1$, we cannot find a value for $\alpha$
that makes both equations coincide with each other.

In this case, let us examine the solution to the equation, i.e., the values for $\Delta r$, $\Delta n$ and $\Delta(d\varpi/dt)$ derived from Eq.~(\ref{s70_2}).
After some calculation we obtain, with  $\beta = \gamma = 1$,
\begin{equation}
\begin{split}
\Delta r &= (2\beta + \gamma - 3\alpha) \frac{\mu}{c^2} =( 3 -  3\alpha) \frac{\mu}{c^2}, \\
\Delta n &= (-4\beta - 2\gamma + 6\alpha) \frac{\mu^2 n}{c^2 h^2}
= (-6 + 6\alpha)  \frac{\mu^2 n}{c^2 h^2}, \\
\Delta\frac{d\varpi}{dt} &= (-\beta + 2\gamma + 2)  \frac{\mu^2 n}{c^2 h^2} = 3  \frac{\mu^2 n}{c^2 h^2}.
\end{split}
\label{s70_3}
\end{equation}
The values in $\Delta r$ and $\Delta n$ contain errors of $O(e^1)$
but rigorous as for $\Delta(d\varpi/dt)$, the same as in Section \ref{S60}.
Here we notice that the values of $\Delta r$, $\Delta n$ and $\Delta(d\varpi/dt)$ in Eq.~(\ref{s70_3}) coincide with those we have obtained in Section \ref{S60},
with $\alpha = 8/3$ as to $\Delta r$ and $\Delta n$ and regardless of the value of $\alpha$
as to $\Delta (d\varpi/dt)$.

From the examination in this section it may be said that the equation of motion
in our study and in PN/PPN approximation seems to be quite different
but the solution, which is the relativistic motion itself of the planet, is not so largely different.

Regarding the difference between Eq.~(\ref{s70_2}) and Eq.~(\ref{s61_13}) and so on,
the source of the mismatch is likely to lie in the differences of the time and length adopted in both theories
but we will leave its solution to the next study.

\section{Concluding remarks}
\label{S80}

Calculations have been made for the relativistic motion of a planet referred to the sun and that of the sun referred to the planet independently
and from the comparison of both the solutions it has been concluded that the spatial and time scales in S-system fixed to the sun and in P-system fixed to the planet must be different.

The sun's motion in P-system is the same as the motion in the Newtonian mechanics, while the planet's motion in S-system appears to be perturbed by the relativistic effect.
However, the apparent perturbation in the planet's motion is caused by the scale difference
in $\tilde{\textrm{S}}$-system and in P-system
and would disappear if the scale difference is suitably taken into consideration.
It means that there is no actual perturbation caused by the relativistic effect both in the planet's and the sun's motions concerning the process of calculation stated so far.

Meanwhile, the difference in the scales in the two systems brings about a real perturbation in the motions of the planet and the sun. 
The motion of the planet relative to the sun in $\tilde{\textrm{S}}$-system,
in which the difference of the spatial and time scales from those in the inertial system is taken into account, 
is subjected to the perturbations in the radius vector $r$, the mean motion $n$ 
and the precession of perihelion $\Delta (d\varpi/dt)$ of the orbital elements.

The equation of motion and its solution thus obtained are expected to be coincident to
those by the general relativity in PN or PPN approximation.
The result of the comparison shows that the equations of motion of ours and of PN/PPN seem to be largely different,
but the values of $\Delta r$, $\Delta n$ and $\Delta (d\varpi/dt)$ coincide with each other
for a certain value of the value of the parameter $\alpha$ as to $\Delta r$ and $\Delta n$
and independently of the value of $\alpha$ as to $\Delta (d\varpi/dt)$.

Now, the difference in the values of  $\Delta r$ and $\Delta n$ may not be so significant.
They decide the difference of the spacetime concerned from another spacetime but it is not certain whichever is the absolute one.
What we can say is only that the spacetime we adopt in the present study is the one we define as
$\tilde{\textrm{S}}$-system and it is different from S-system
which we suppose to be an inertial system but not very definitely.

On the other hand, $\Delta (d\varpi/dt)$ is a physical quantity with a definite meaning.
The quantity $d\varpi/dt$ is the ratio of the angle the planet's perihelion gains
to that the planet advances on its orbit during the same time interval.
This ratio would not depend on the kind of the adopted spacetime.
The fact that the expression for $\Delta (d\varpi/dt$) in Eq.~(\ref{s70_3}) does not contain $\alpha$ may suggest it.

Further, one more fact is noticed as to $\Delta (d\varpi/dt)$.  As far as the present study is admissible,
the root of the relativistic perihelion precession of a planet lies in Eq.~(\ref{s62_1}),
which also stems from the difference of the scales between in $\tilde{\textrm{S}}$-system and S-system  given in Eq.~(\ref{s5_3}). 

Finally, it would be significant to inquire what spatial and time scales we have on the earth, using which the astronomical observation is made.
We, who move together with the earth, are not in $\tilde{\textrm{S}}$-system but in P-system
which is making a free fall except for a negligibly small effect owing to the difference of the positions between the center and the surface of the earth.
On this point the spatial and time scales that we recognize are those in the inertial system, i.e., in I-system.

However, it must be noted that there exists the earth's gravity on its surface, which is considerably strong as compared with the gravitation owing to the sun at that place.
The rate of both effects on the scales is roughly $(E/M)(a/a_e) \cong 0.05$ with $M$ and $E$ being the masses of the sun and the earth, respectively,
and $a$ and $a_e$ the sun-earth distance and the earth's radius, respectively.  
Accordingly, our scales are far from those in the inertial system and this fact must be taken into consideration if necessary.
Only on an artificial satellite in whatever orbit, the scales in the inertial system would be realized.

\section*{Acknowledgments}

The auther sincerely thanks Dr. Michael Efroimsky of US Naval Observatory for his valuable comments and strong support.

\end{document}